\documentclass[aps, pra, reprint, longbibliography, 
superscriptaddress]{revtex4-1}

\usepackage{hyperref}
\usepackage{graphicx}
\usepackage[utf8]{inputenc}
\usepackage{xcolor}
\usepackage{amsmath, amssymb}
\usepackage{physics}
\usepackage{ulem}

\begin{document}



\title{Ring solids and supersolids in spherical shell-shaped dipolar 
Bose-Einstein condensates}

\author{J. S\'anchez-Baena}
\email{juan.sanchez.baena@upc.edu}
\affiliation{Departament de F\'isica, Universitat Polit\`ecnica de Catalunya, Campus Nord B4-B5, 08034 Barcelona, Spain}
\author{R. Bombín}
\affiliation{Departament de F\'isica, Universitat Polit\`ecnica de Catalunya, Campus Nord B4-B5, 08034 Barcelona, Spain}
\affiliation{Université de Bordeaux, 351 Cours de la Libération, 33405 Talence, France}
\author{J. Boronat}
\affiliation{Departament de F\'isica, Universitat Polit\`ecnica de Catalunya, Campus Nord B4-B5, 08034 Barcelona, Spain}

\date{\today}

\begin{abstract}
We study the interplay between the anisotropy of the dipole-dipole interaction 
and confinement in a curved geometry by means of the extended Gross-Pitaevskii equation, which allows us to characterize the ground state of a dipolar Bose gas under the confinement of a bubble trapping potential. We do so in terms of the scattering length $a$ and the number of particles. 
We observe the emergence of a wide variety of dipolar solids, consisting on arrangements of different number of droplets along a ring over the equator of the spherical shell confinement.
We also show that the transition between the different phases of the 
system can be engineered by varying $a$, the number of particles or the radius 
of the trap, parameters which can be experimentally tuned. 
Finally, we show the importance of working in microgravity conditions as gravity 
unstabilizes the observed dipolar solids.
\end{abstract}

\maketitle

\section{\label{sec:introduction}Introduction}

Since the realization of the first Bose-Einstein condensate (BEC) in 1995, the 
research on ultracold gases has been boosted by the achievement of a  
high control over them. This allows to explore not only a wide range 
of interaction strengths 
but also different geometries, ranging from one to three dimensions.
In recent years, with the development of trapping techniques, it has been 
possible
to achieve more exotic geometries with non-trivial topologies 
\textit{eg.}, 
rings or curved surfaces (see Ref.~\cite{Garraway2016} for a review on trapping 
techniques).
These advances have motivated experiments both in zero gravity 
conditions~\cite{Becker2018} and with a gravity compensation 
mechanism~\cite{Guo2022} 
where the gravitational sag is absent and the BEC
can be engineered in the shell of a sphere~\cite{Carollo2022} (see~\cite{tononi2023} for recent reviews on the topic).
Previous works include the characterization of BEC condensation and 
excitations~\cite{Moller2020,Tononi2020}, the topological superfluid phase
transition~\cite{Tononi2019,Tononi2022}, 
the BEC-BCS crossover~\cite{He2022}, the gas to soliton 
transition~\cite{tononi2023c}, the study of vortices and collective
excitations~\cite{tononi2023b} and the application of matter-wave lensing
techniques~\cite{Boegel2023}.
On a less fundamental approach, the possibility of engineering atom-based 
circuits has also been explored~\cite{Amico2022}.

In the context of ultracold gases, the study of dipolar systems has revealed astonishing phenomena such as droplet formation
and the emergence of supersolidity~(see Ref.~\cite{Chomaz2023} for an 
experimental  review). Supersolidity refers to a state of matter 
that simultaneously features spatial diagonal and off-diagonal long-range order 
~\cite{Andreev:JETP:1969,Chester:PRA:1970,Leggett:PRL:1970}. In fact, dipolar 
systems 
emerge as an exceptional setup for  
studying the phenomena of supersolidity. This topic has been extensively 
studied~\cite{Bottcher2021,Buchler2022,Hertkorn2021,Gallemi2022,
Hertkorn2021_2, 
Blakie2020,Bombin2017,Bombin2019,Zhang2019,Zhang2021,Zhang2023,Blakie2020_2, 
Smith2023,Baena2023} and experimentally 
confirmed~\cite{Guo2019,Tanzi2019B,Natale2019,Tanzi2021,Petter2021,Biagioni2022, 
Bland2022,Norcia2022,Tanzi2019,Bottcher2019,Chomaz2019,Norcia2021}.
Supersolid phases have been also predicted to occur in strictly two-dimensional 
geometries, where the transition to the normal state is of the 
Berezinskii-Kosterlitz-Thouless type~\cite{Filinov2010,Bombin2017,Bombin2019,Ota2018}.  Nonetheless, 
only a few studies have considered this phenomena
on curved surfaces (see Refs.~\cite{Diniz2020,Ciardi2024,Arazo2021,Tengstrand2023,Young2023,Sindik2023,Mukherjee2024, ghosh2024}).

 \begin{figure*}[t]
\centering
\includegraphics[width=1\linewidth]{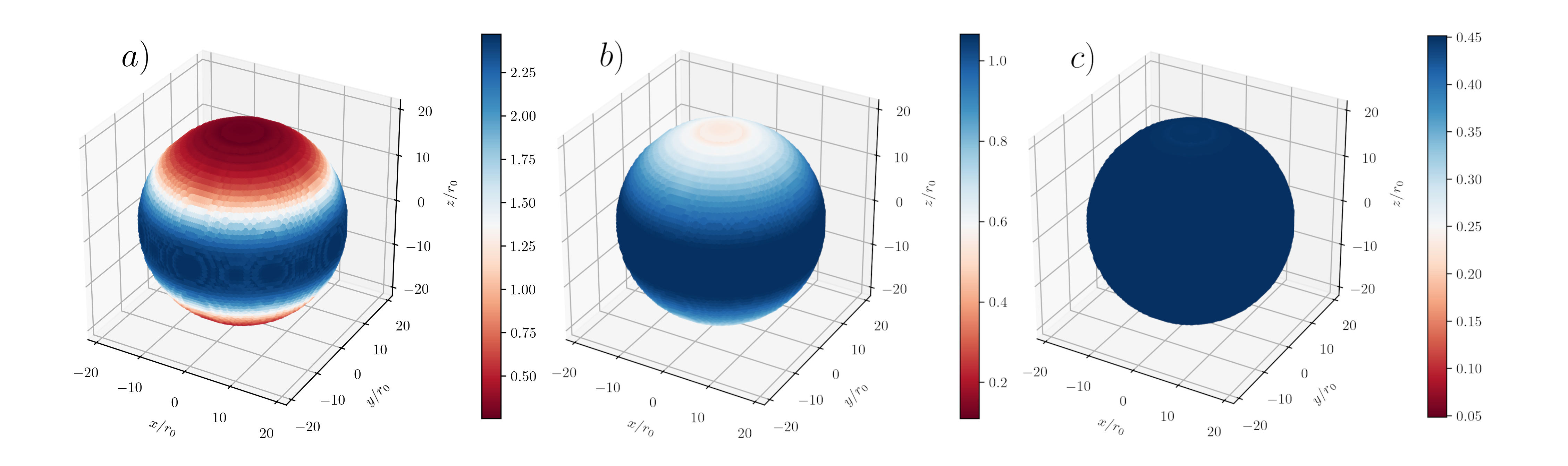}
\caption{Three-dimensional probability density of the dipolar 
BEC under a shell-shaped confinement (see Eq.~\ref{eqn:bubble_trap}) 
 with detuning $\Omega/\epsilon = \Delta/\epsilon = 400$ and bare harmonic 
frequency $\omega_0 = 0.22 \epsilon/\hbar$. The probability density is reported 
at the surface $r = \sqrt{x^2 + y^2 + z^2} = r_0 \sqrt{\Delta/\epsilon}$ 
for $\varepsilon_{\text{dd}} = 1.25 \text{ }(a), 0.5\text{ }(b)$ and $0.1 
\text{ }(c)$. The colorbars indicate the value of $\abs{\Psi({\bf r})}^2$ in 
units of $r_0^{-3}$. }
\label{fig:fig1}
\end{figure*}

Dipolar shell-shaped systems are expected to exhibit a richer phase diagram than 
contact BEC gases~\cite{tononi2023}. On one hand, the long ranged and 
anisotropic character of dipolar interaction is known to produce density 
modulated phases with important long-range correlations in free space. 
On the other hand, the curvature would make the ground state of the system very 
different to that of the free space, for example by frustrating the formation of 
stripes. In this sense, the interplay between anisotropy, long-range order, 
and topology in spherically symmetric traps can give rise to BEC states in which 
the spherical symmetry of the trap is spontaneously broken.

To give some insight into the previously mentioned phenomena, in the present 
work we study a dipolar Bose gas confined on a spherical bubble 
trapping potential in the regime of parameters where supersolidity arises. The 
paper is organized as follows. In Sec.~\ref{sec:theory}, we illustrate the 
methodology that we employ.  The main results of our work are presented and discussed in
Sec.~\ref{sec:results} . Finally, in Sec.~\ref{sec:conclusions} we summarize
the main conclusions and discuss future perspectives. 

\section{\label{sec:theory}Theory}

 We consider a system of $N$ magnetic dipolar atoms of mass $m$ with all their magnetic moments $\mu$ aligned along the $z$-axis. The system is confined in a bubble trap potential~\cite{colombe_2004,sun18}
\begin{align}
 V_{\text{trap}}({\bf r}) &= m \omega_0^2 r_0^2 \sqrt{ \frac{[(r/r_0)^2 - \Delta/\epsilon]^2}{4} + (\Omega/\epsilon)^2 },
 \label{eqn:bubble_trap}
\end{align}
where $\omega_0$ is the frequency of the bare harmonic trap, prior to 
radiofrequency (rf) dressing, and the parameters $\Delta$ and $\Omega$ are the 
detuning between the radiofrequency (rf) field and the different energy states 
employed to prepare the condensate, and the Rabi coupling between these states, 
respectively~\cite{Carollo2022,sun18}. We have also introduced the relevant
length and energy scales, $r_0$ and $\epsilon$ respectively, given 
by $r_0 = 12 \pi a_{\text{dd}}$ and $\epsilon = \hbar^2/(m r_0^2)$, where 
$a_{\mathrm{dd}} =  \frac{C_{\rm dd} m}{12\pi\hbar^2}$ is the dipole length,
with 
$C_{\rm dd} = \mu_0\mu^2$, $\mu_0$ the Bohr magneton and $\mu$ the magnetic dipole
moment of the atoms. For simplicity, and analogously to a previous 
work~\cite{sun18}, we consider $\Delta = \Omega$.

In order to characterize the ground state of the system in a spherically symmetric trap we solve the three-dimensional extended Gross-Pitaevskii equation that reads
\begin{align}
 \mu \Psi({\bf r}) = &\left[-\frac{\hbar^2}{2m} \nabla^2  + V_{\textrm{trap}}({\bf r}) + g | \Psi(r)|^2 + \gamma_{\text{QF}} \abs{\Psi({\bf r})}^3 \right. \nonumber \\
 &\left. + \int d{{\bf r}^\prime} V_{\rm dd}({\bf r} - {\bf r'}) \abs{\Psi({\bf r'})}^2 \right]\Psi({\bf r})  \ ,
 \label{eqn:eGPE}
\end{align}
with $\mu$ the chemical potential, 
$\Psi({\bf r})$ the condensate wave function,  which is normalized as $N = \int 
d{\bf r} \abs{\Psi({\bf r})}^2$, and $g = 4\pi \hbar^2 a_s / m$  the  coupling
constant with $a_\mathrm{s}$ the s-wave scattering length. The fourth term  
$\gamma_{\mathrm{QF}}|\Psi|^3$ is the LHY (Lee-Huang-Yang) correction~\cite{Lima2011,Lima2012}
which introduces quantum fluctuations,
\begin{equation}
\gamma_{\mathrm{QF}}|\Psi|^3=\frac{32 g \sqrt{a_s^3}}{3 \sqrt{\pi}} \mathcal{Q}_{5}\left(\varepsilon_{\mathrm{dd}}\right)|\Psi|^3,
\label{eqn:LHY}
\end{equation}
with $\varepsilon_{\mathrm{dd}} = a_{\mathrm{dd}} / a_s$ and 
$Q_5(\varepsilon_{\mathrm{dd}}) = \dfrac{1}{2} \displaystyle \int_{0}^{\pi} 
\mathrm{d} \alpha \sin\alpha \left[1 + \varepsilon_{\rm dd} (3 \cos^2 \alpha - 
1)\right]^{5/2}$. 
Finally, the last term in Eq.~\eqref{eqn:eGPE} accounts for the dipole-dipole 
interaction (DDI), 
\begin{align}
V_{\rm dd}({\bf r} - {\bf r'}) &= \frac{C_{\text{dd}}}{4 \pi} \frac{1 - 3\cos^2 \theta}{\abs{{\bf r} - {\bf r'}}^3},
\label{pseudopot}
\end{align}
 where $\theta$ is the polar angle of the vector ${\bf r} - {\bf r'}$.

The use of the pseudopotential in Eq.~\ref{eqn:eGPE} (that is, the term $g | \Psi(r)|^2 + \int d{{\bf r}^\prime} V_{\rm dd}({\bf r} - {\bf r'}) \abs{\Psi({\bf r'})}^2$) is justified as long as the
collisions between particles can be treated as three-dimensional processes. The 
system lays in the two dimensional 
regime if the harmonic length $a_{\text ho} = \sqrt{\hbar/(m \omega_{0})}$, 
which is associated to the tightness of the confinement, is significantly 
smaller than any other length scales. However, in our calculations, we choose a 
trapping strength such that $ a_{\text ho} \gg a_\mathrm{s}$. Therefore, all 
scattering processes can be considered three-dimensional and the pseudopotential 
of Eq.~\ref{eqn:eGPE} can be applied. In the case of a tight confinement (thin shell limit), a
pseudopotential that considers the effects of the geometry should be 
employed~\cite{tononi22scattering,zhang18}.

In the majority of this work, we restrict ourselves to the zero gravity limit. However, the effect of a gravitational force can be accounted for by adding the following one-body potential~\cite{Arazo2021} to Eq.~\ref{eqn:eGPE}
\begin{align}
 V_g({\bf r}) = mg(x \sin \theta_g + z \cos \theta_g) \ ,
 \label{eqn:gravity}
\end{align}
where $\theta_g$ is the angle between the $z$-axis and the gravity
direction. For $^{164}$Dy atoms, the gravitational strength on the Earth 
corresponds to to $mg = 1.15 \epsilon/r_0 = m g_{\rm E}$. In 
Sec.~\ref{sec:gravity}, we study the robustness of a dipolar solid ring under 
the effect of gravity.

\section{\label{sec:results}Ring solids and supersolids}

As a means to illustrate the system under study, we show in Fig.~\ref{fig:fig1} the probability density of the dipolar BEC confined within the bubble trap under zero gravity for different values of the ratio $\varepsilon_{\mathrm{dd}} = a_{\mathrm{dd}} / a_s$. We have employed a trap with parameters
$\Delta/\epsilon=400$, $\omega_0 = 0.22 \epsilon/\hbar$, which correspond to
$\omega_0 = 2\pi \times 200$ Hz and a bubble trap with radius $R=
r_0 \sqrt{\Delta/\epsilon}=5.2$ $\mu$m for $^{164}$Dy atoms. These parameters are realistic
for a setting with non-dipolar gases. From the figure, we can see that in the 
contact dominated regime ($\varepsilon_{\mathrm{dd}} \ll 1$), the dipolar gas 
fills up the spherical shell, yielding an apparently spherically symmetric 
density distribution, despite the anisotropy of the DDI. 
As seen in Fig.~\ref{fig:fig1}, and as reported in previous 
works~\cite{Arazo2021,Ciardi2024}, the atoms of the BEC gas tend to populate 
the equator of the shell, even before reaching the dipole dominated regime, 
$\varepsilon_{\mathrm{dd}}>1$. This magnetostriction is a consequence of the 
competition between the anisotropy of the dipole-dipole interaction (which 
energetically favours head-to-tail arrangements of dipoles) and the shell shape 
of the external confinement. Experimentally, ring shaped condensates can be 
obtained by the use of toroidal tarps~\cite{Javanainen1998, 
Tengstrand2021,Tengstrand2023,Zhang2022}, which can also be realized in 
experiments~\cite{Ramanathan2011,Wright2013,Wright2013_2}. However, in the 
present case, as well as for cylindrically-shaped traps~\cite{Roccuzzo2022}, the 
ring shape of the condensate arises purely due to magnetostriction in a 
spontaneous manner, instead of being entirely forced by the external 
confinement. For the rest of our work, we focus on the regime 
$\varepsilon_{\mathrm{dd}}>1$ and the aforementioned value of the trapping 
strength.

 \begin{figure}[t]
\centering
\includegraphics[width=1.00\linewidth]{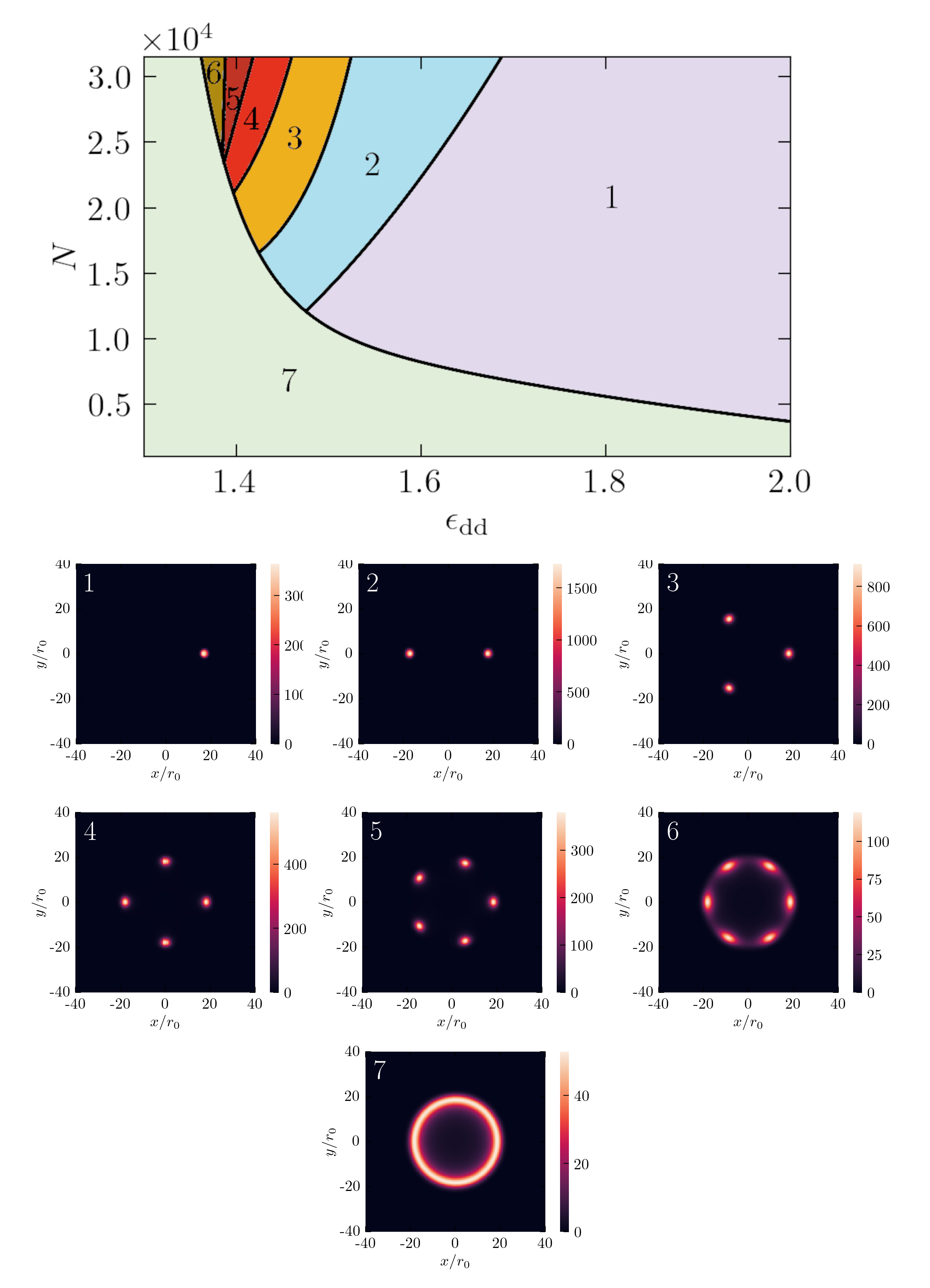}
\caption{ Top: structural diagram of the ring-shaped dipolar condensate as a function of $\epsilon_{\rm dd} = a_{\rm dd}/a$ and the number of particles $N$. Panels 1-7: integrated density profiles $\rho(x,y)$ of the condensate density arising in each region of the diagram. The colorbars indicate the value of $\rho(x,y)$ in units of $r_0^{-2}$. The parameters for the bubble trap (see Eq.~\ref{eqn:bubble_trap}) are $\Delta/\epsilon=\Omega/\epsilon=400$, $\omega_0 = 0.22 \epsilon/\hbar$.} 
\label{fig:fig2}
\end{figure}

In the dipolar dominated regime ($\epsilon_{\rm dd} > 1$), the anisotropy of the 
dipolar interaction can give rise
to dipolar solids and supersolids, in analogy to the phenomenology that takes 
place in bulk-trapped dipolar BECs. Eq.~\ref{eqn:eGPE} can numerically be solved 
for different 
values of the scattering length $a_\mathrm{s}$ (and thus, $\epsilon_{\rm dd}$) 
and number of particles $N$ to obtain the ground state of the system. As
reported in similar works~\cite{Tengstrand2021,Baillie2018,Roccuzzo2020}, the 
energy minimization has to be carefully performed, sampling a wide variety of 
initial conditions, as many metastable states close to the ground state 
exist. We detail in the Appendix~\ref{sec:numerics} the numerical parameters of 
our simulations, as well as the initial conditions considered. Since superfluid 
structures
may arise,
we evaluate the Leggett's upper bound estimator for 
superfluidity~\cite{Leggett1998}, which is computed from the ground state 
density as
\begin{align}
 f_s = \left[ \frac{1}{2 \pi} \int_0^{2 \pi} \frac{d\theta}{\rho(\theta)/\rho_0} \right]^{-1} \ ,
 \label{eqn:legget}
\end{align}
where
\begin{align}
 \rho(\theta) &= \int dz dr \text{ }r \abs{\Psi(r,\theta,z)}^2,
 \\
 \rho_0 &= \frac{1}{2 \pi} \int_0^{2 \pi} \rho(\theta) d\theta \ .
\end{align}
Eq.~\ref{eqn:legget} yields the trivial unity limit for a ring-like condensate 
(since $\rho(\theta)$ is a constant) and decreases as the modulation of the 
wave function increases along the ring. This upper-bound 
estimator for the superfluidity has shown excellent agreement with the  
calculation of non-classical translational inertia for a system of dipoles 
confined in a quasi-1D tubular geometry~\cite{Smith2023}.

\subsection{Structural diagram and transitions}

The structural diagram of the system is reported in Fig.~\ref{fig:fig2}, where 
we show the two-dimensional integrated density $\rho(x,y)=\int dz \abs{\Psi({\bf 
r})}^2$ of each structure. Regarding the superfluid fraction along the ring, 
only the structures featuring dipolar clusters in the interval $\varepsilon_{\rm 
dd} \in (1.36, 1.41)$ yield a significant superfluidity ($f_s > 0.2$), while 
for $\varepsilon_{\rm dd} > 1.41$, the result of Eq.~\ref{eqn:legget} quickly 
drops to zero. On the other hand, all the states in the SF region of 
Fig.~\ref{fig:fig2} yield unit superfluidity. We label the states with $f_s = 1$ 
as superfluids, while we call supersolids and solids those which yield $f_s > 
0.2$ and $f_s < 0.2$, respectively.
In general, the increase of $\epsilon_{\rm dd}$ for a fixed number of particles 
implies 
a lower number of droplets, since the attraction of the DDI favours the bunching of dipoles and thus, fewer and more elongated clusters are
formed. In much the same way, the decrease of the number of particles for a 
fixed $\epsilon_{\rm dd}$ also causes a reduction in the number of droplets 
because the system wants to maximize the number of dipoles 
placed in a head-to-tail configuration. 
The peak density of the droplets that we obtain lies close to the expected 
values obtainable with harmonic traps (see the seminal experiment of 
Ref.~\cite{Kadau2016}). For the largest number of particles ($N=31500$) the peak 
density lies in the interval $\rho_{\rm peak} \in [3, 100] \times 10^{14}$ 
cm$^{-3}$, where the largest values are achieved for $\varepsilon_{\rm dd}=2$, 
for which all particles cluster into a single droplet.
We also see that the solid structures 
disappear if the number of particles is decreased below a threshold, from which
there is not enough density to sustain clustering. This phenomenology is 
reminiscent of a quasi-1D system of dipoles confined in a 
tube~\cite{Blakie2020_2,Smith2023} where the disappearance of the supersolid 
phase in the low density regime upon decreasing the density is also reported.

\begin{figure}[t]
\centering
\includegraphics[width=0.9\linewidth]{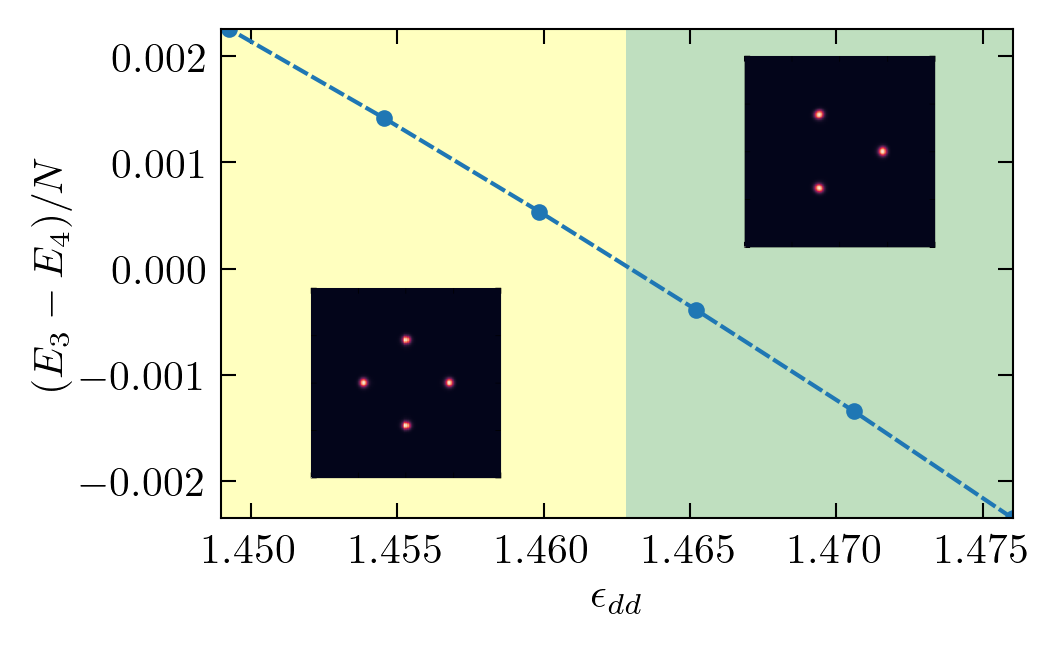}
\caption{ Difference in the energy per particle between a solid state with 3 and 4 droplets as a function of $\varepsilon_{\rm dd}$ for $N=31500$ and the same bubble trap as in Fig.~\ref{fig:fig2}. }
\label{fig:fig3}
\end{figure}

\begin{figure}[t]
\centering
\includegraphics[width=0.9\linewidth]{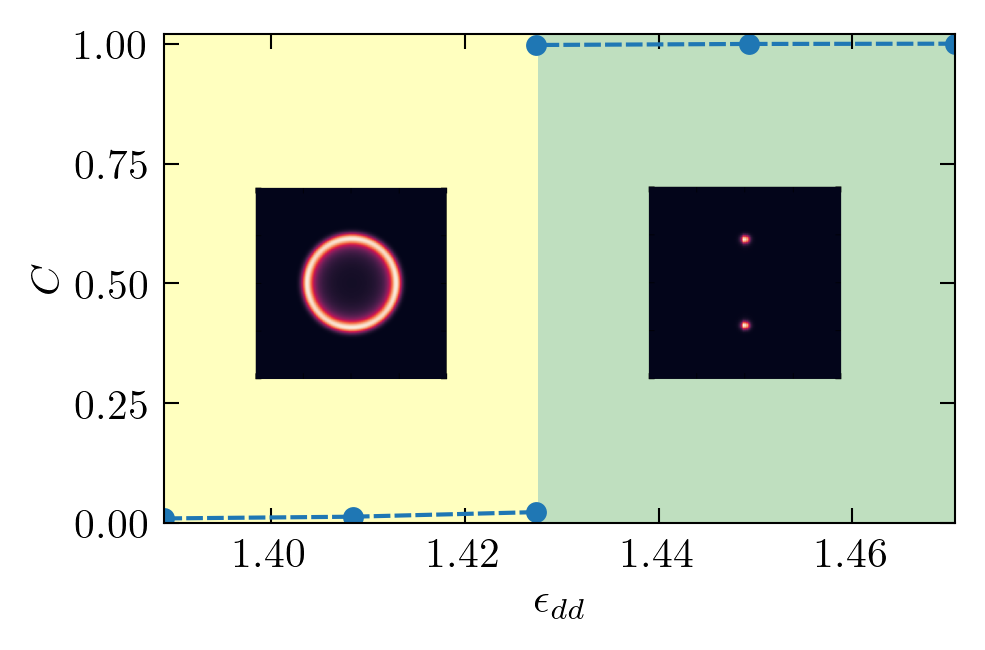}
\caption{ Contrast of the BEC density (see Eq.~\ref{eqn:contrast}) as a function of $\varepsilon_{\rm dd}$ for $N=16500$. The bubble trap parameters are the same as in Fig.~\ref{fig:fig2}. }
\label{fig:fig4}
\end{figure}

Precisely, in the spirit of these works, it is interesting to examine the character (continuous or discontinuous) of the transition between the different structures present in the diagram of Fig.~\ref{fig:fig2}. We can not strictly speak of first and second order phase transitions (as it is done in~\cite{Blakie2020_2,Smith2023}) because our system is finite. Our calculations show that the transition between different solid states is discontinuous, meaning that the system jumps from a state with a given number of droplets to a different one discontinuously (the density distribution changes abruptly). This is because there exists an energy crossing between the different metastable states at the transition boundary. We illustrate this in Fig.~\ref{fig:fig3}, where we report the energy difference between two dipolar solids across the transition between regions 3 and 4 of the diagram in Fig.~\ref{fig:fig2} for $N=31500$. In much the same way, the transition between a dipolar solid and the superfluid is also discontinuous, in analogy to the first order phase transition that takes place in the low density regime of the tubular quasi-1D system. This is illustrated in Fig.~\ref{fig:fig4}, where we show the contrast of the BEC density across the transition between regions 2 and 7 of Fig.~\ref{fig:fig2}, for $N=16500$. Here, the contrast is defined as
\begin{align}
 C = \frac{\rho_{\rm max.} - \rho_{\rm min.}}{\rho_{\rm max.} + \rho_{\rm min.}}
 \label{eqn:contrast}
\end{align}
where $\rho_{\rm max.}$ and $\rho_{\rm min.}$ are, respectively, the maximum and minimum values of the integrated density $\rho(x,y)$ along the circle of radius $R/r_0 = \sqrt{\Delta / \epsilon}$. As one can see from Fig.~\ref{fig:fig4}, the contrast in the density shows a clear discontinuity. 
Therefore, in view of the results, we can draw similarities between the physics 
under our geometry and the quasi-1D tubular one, since the transition that leads 
to a superfluid gas in our system is analogous to the first-order 
phase transition of Refs.~\cite{Blakie2020_2,Smith2023}. 

\begin{figure}[t]
\centering
\includegraphics[width=0.9\linewidth]{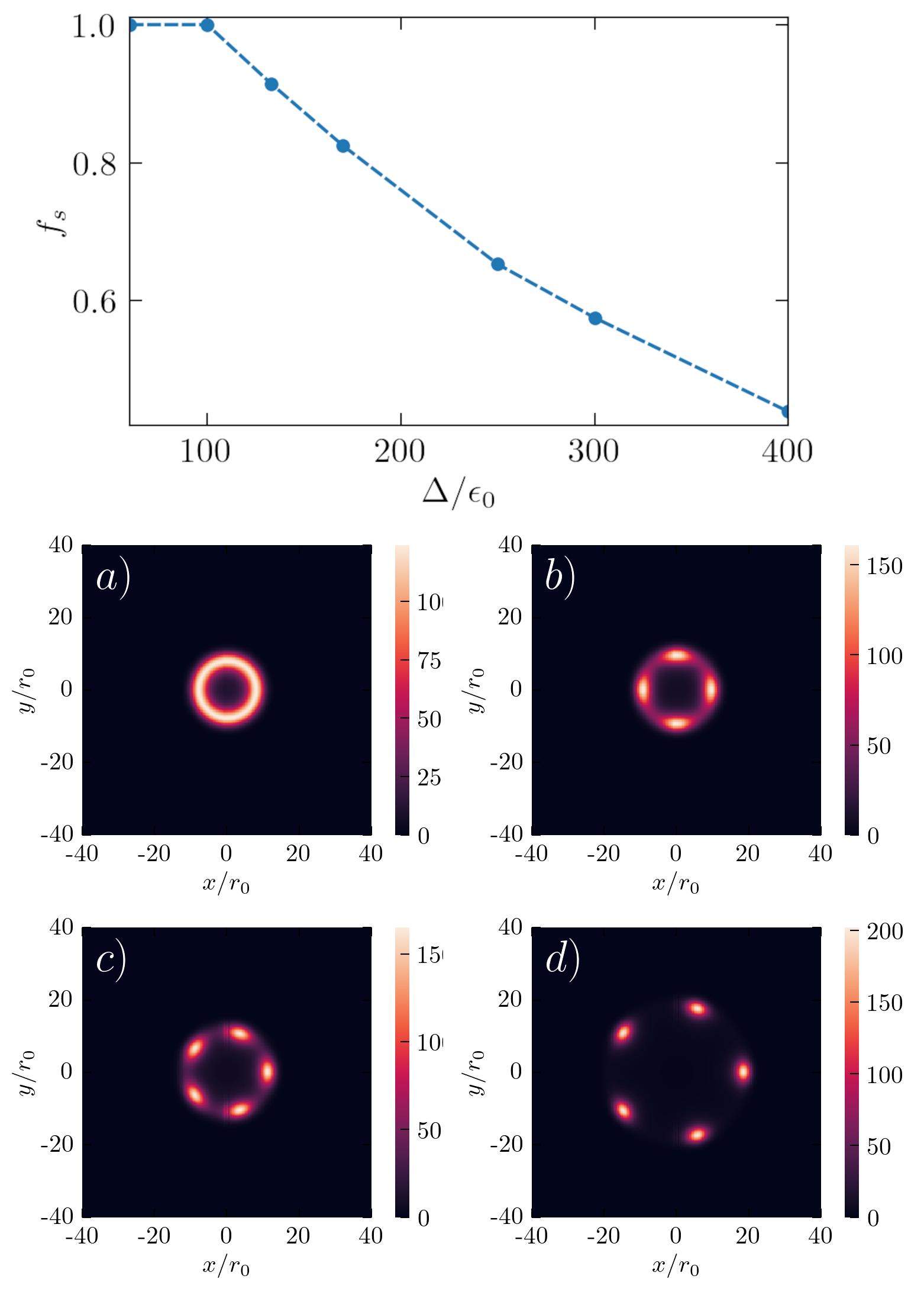}
\caption{ \textit{Top:} Superfluid density as a function of the detuning
$\Delta/\epsilon$. \textit{Bottom:}
Integrated density $\rho(x,y)$ for detuning
values $\Delta/\epsilon=100, 133, 170$, and $400$ from a) to d),
respectively. The colorbars indicate the value of $\rho(x,y)$ in units of $r_0^{-2}$. The ratio between the dipole length and the scattering length is
set to $\epsilon_{\rm dd} = 1.38$ and  $N=26500$. The
bare harmonic frequency is the same as in Fig.~\ref{fig:fig2}. }
\label{fig:fig6}
\end{figure}

\begin{figure}[t]
\centering
\includegraphics[width=0.9\linewidth]{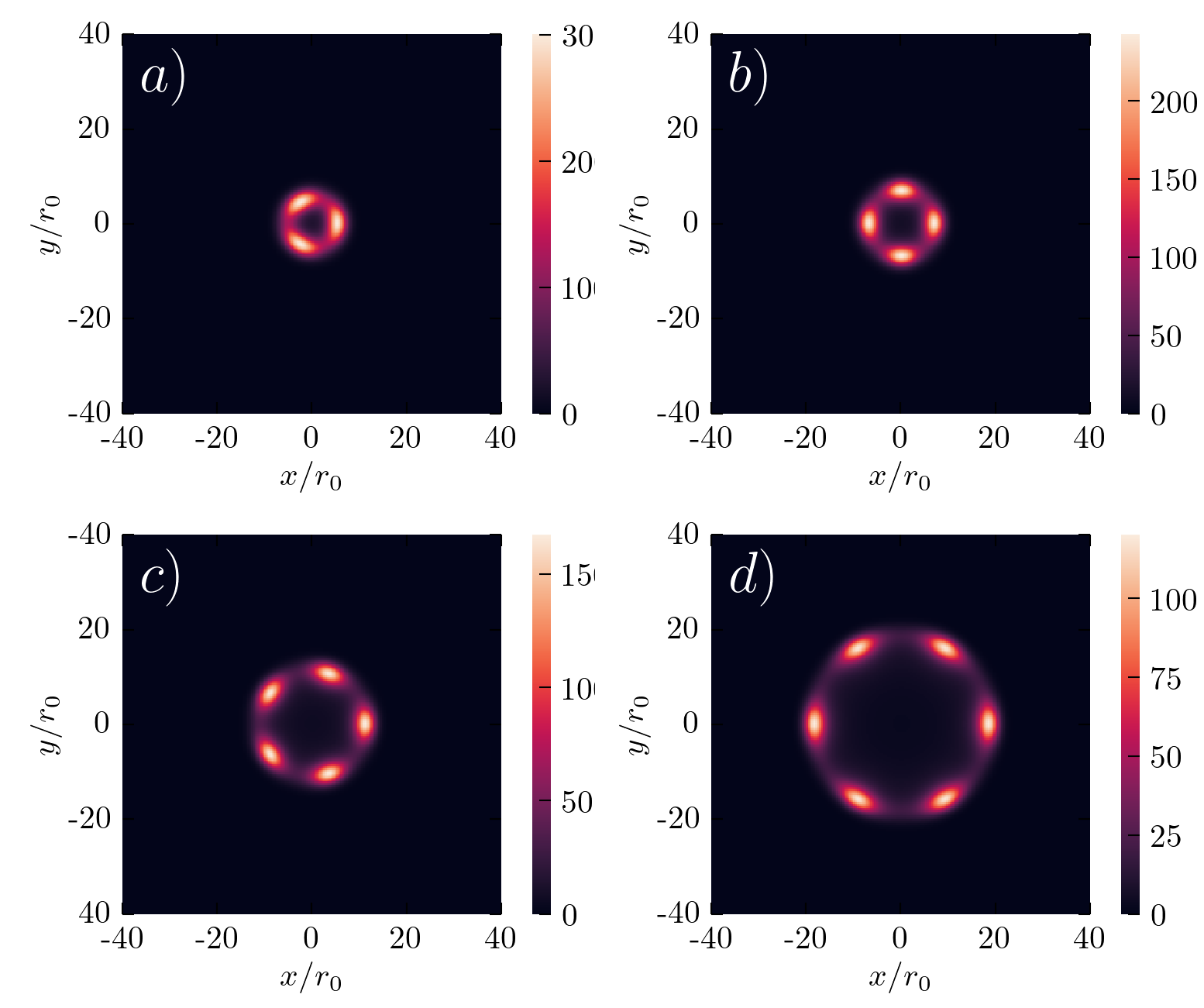}
\caption{Ring shaped supersolid dipolar states are engineered by appropriate combinations of values of the magnetic field detuning, the scattering length and the number of particles. Panels a)-d) show the integrated density $\rho(x,y)$ of supersolid states obtained for $\epsilon_{\rm dd}=1.43$, $N=26500$, $\Delta/\epsilon=50$ (a), $\epsilon_{\rm dd}=1.41$, $N=26500$, $\Delta/\epsilon=80$ (b), $\epsilon_{\rm dd}=1.39$, $N=26500$, $\Delta/\epsilon=170$ (c), $\epsilon_{\rm dd}=1.37$, $N=31500$, $\Delta/\epsilon=400$ (d). The colorbars indicate the value of $\rho(x,y)$ in units of $r_0^{-2}$. The bare harmonic frequency is the same as in Fig.~\ref{fig:fig2}. }
\label{fig:fig7}
\end{figure}

\subsection{Engineering supersolids}

Even though the supersolid phase constitutes a small region in the diagram of
Fig.~\ref{fig:fig2} as mentioned previously, a rich variety of supersolid structures can
be engineered when tuning $\Delta$, and thus effectively modifying the radius of
the trapping spherical shell. By increasing $\Delta$ starting from a conventional harmonically trap gas, the system transitions to a supersolid. Furtherly increasing $\Delta$ leads to an increase in the number of droplets, all while retaining a substantial superfluid fraction and the
ring shape. This is shown in
Fig.~\ref{fig:fig6}, where we report the integrated density $\rho(x,y)$ of the
condensate wave function for different values of $\Delta$. We also report
Leggett's upper bound for the superfluid fraction (see Eq.~\ref{eqn:legget}),
which increases as the trap radius decreases, thus confirming that the
superfluid density can be enhanced by reducing $\Delta$.
The variation of this parameter is pretty straightforward in experimental
setups, as changing $\Delta$ is precisely how the bubble trap is generated from
a harmonic potential. The wide variety of solid structures in the diagram of
Fig.~\ref{fig:fig2} thus allows for the observation of many
different supersolid dipolar rings (i.e. supersolids with a different number of droplets) by playing with the parameter $\Delta$ starting from different points of the structural diagram. This is showcased in Fig.~\ref{fig:fig7}, where we show four examples of different supersolids obtained for
different combinations of the number of particles, the detuning and the
scattering length.

\subsection{High particle number limit}

\begin{figure*}[t]
\centering
\includegraphics[width=0.9\linewidth]{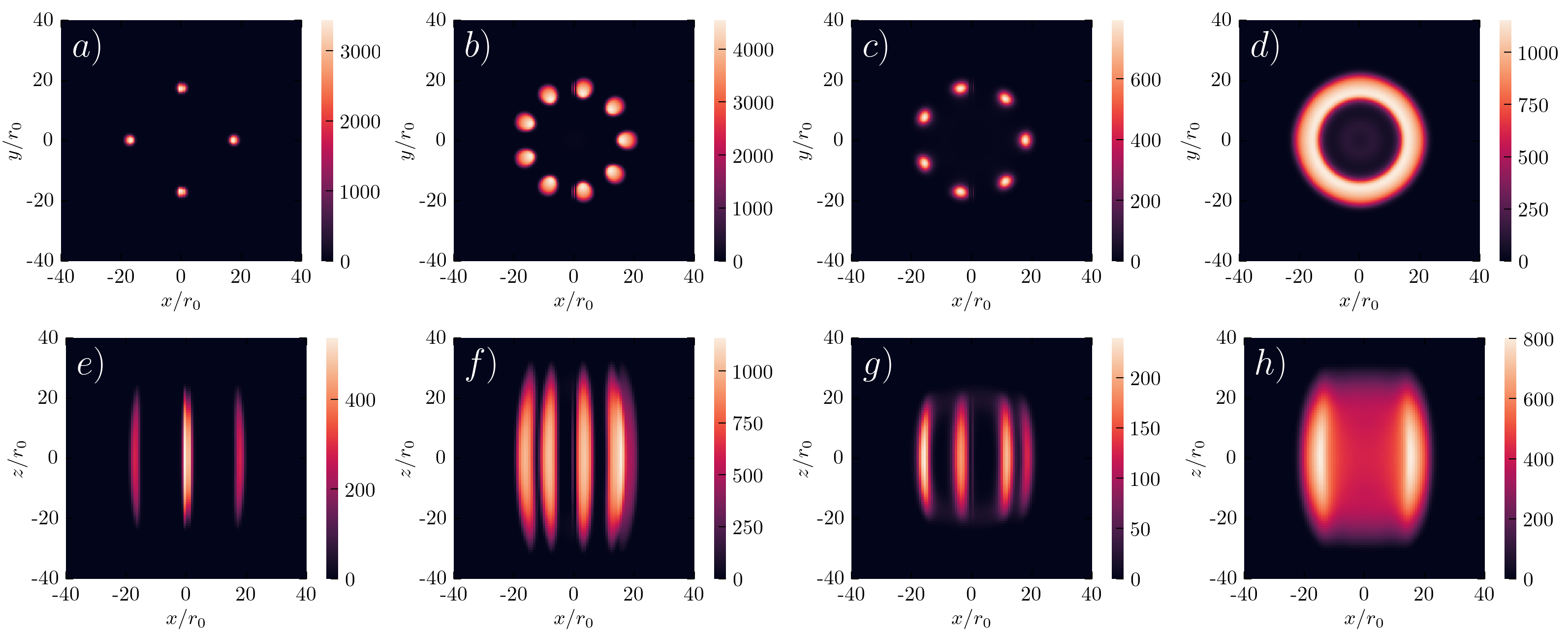}
\caption{ Integrated densities $\rho(x,y)$ (top) and $\rho(x,z)$ (bottom) for $\varepsilon_{\rm dd} =1.72$, $N=10^5$ ((a), (e)), $\varepsilon_{\rm dd} =1.72$, $N=10^6$ ((b), (f)), $\varepsilon_{\rm dd} =1.42$, $N=10^5$ ((c), (g)) and $\varepsilon_{\rm dd} =1.42$, $N=10^6$ ((d), (h)). The bubble trap parameters are the same as in Fig.~\ref{fig:fig2}. The colorbars indicate the value of $\rho(x,y)$ and $\rho(x,z)$ in units of $r_0^{-2}$.  }
\label{fig:fig5}
\end{figure*}

Up to now, we have restricted the particle number to the interval $N \leq 31500$. However, it is interesting to consider higher particle numbers, specifically to check whether a secondary, or even multiple rings can arise. In order to address this question, we have computed the ground state of the dipolar BEC for $N = 10^5$ and $N = 10^6$ for two different values of $\varepsilon_{\rm dd}$ ($\varepsilon_{\rm dd} =1.42$ and $\varepsilon_{\rm dd} =1.72$). For these cases, we show in Fig.~\ref{fig:fig5} the integrated
two-dimensional densities $\rho(x,y)=\int dz \abs{\Psi({\bf r})}^2$ and $\rho(x,z)=\int dy \abs{\Psi({\bf r})}^2$ . As we can
see from the figure, atoms accumulate on the central ring instead of forming 
additional ones, which gets wider as the number of particles increases. Looking 
at the case with $\varepsilon_{\rm dd} =1.72$, one can also see that increasing 
the number of particles leads the system to an unstructured superfluid, meaning 
that there exists an upper threshold for the particle number above which the 
solid and supersolid structures disappear. Again, this is in line to the 
phenomenology reported in the quasi-1D tubular geometry, where in the high 
density region of the phase diagram, increasing the density drives a 
supersolid-to-superfluid phase transition~\cite{Blakie2020_2,Smith2023}.

\subsection{\label{sec:gravity}Effect of gravity}

\begin{figure}[t]
\centering
\includegraphics[width=1.0\linewidth]{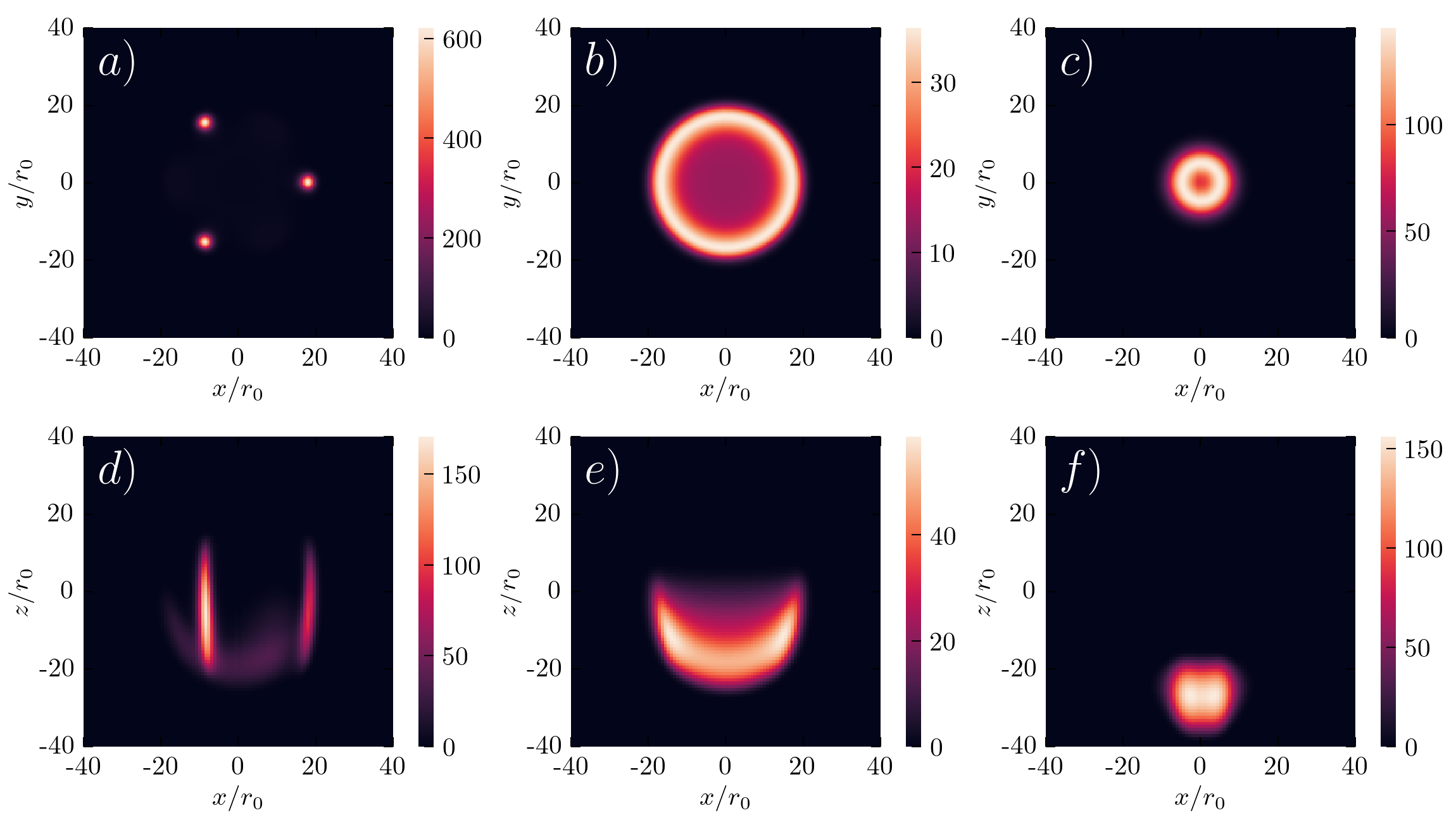}
\caption{Integrated densities $\rho(x,y)$ ((a)-(c)) and $\rho(x,z)$ ((d)-(f)) for a varying gravitational strength $mg = 0.03 m g_{\rm E}$ ((a), (d)), $mg = 0.04 m g_{\rm E}$ ((b), (e)) and $mg = 0.5 m g_{\rm E}$ ((c), (f)) and an angle $\theta_g = 0$ (see Eq.~\ref{eqn:gravity}). The bubble trap parameters are the same as in Fig.~\ref{fig:fig2}. The colorbars indicate the value of $\rho(x,y)$ and $\rho(x,z)$ in units of $r_0^{-2}$. }
\label{fig:fig8}
\end{figure}

\begin{figure}[t]
\centering
\includegraphics[width=1.0\linewidth]{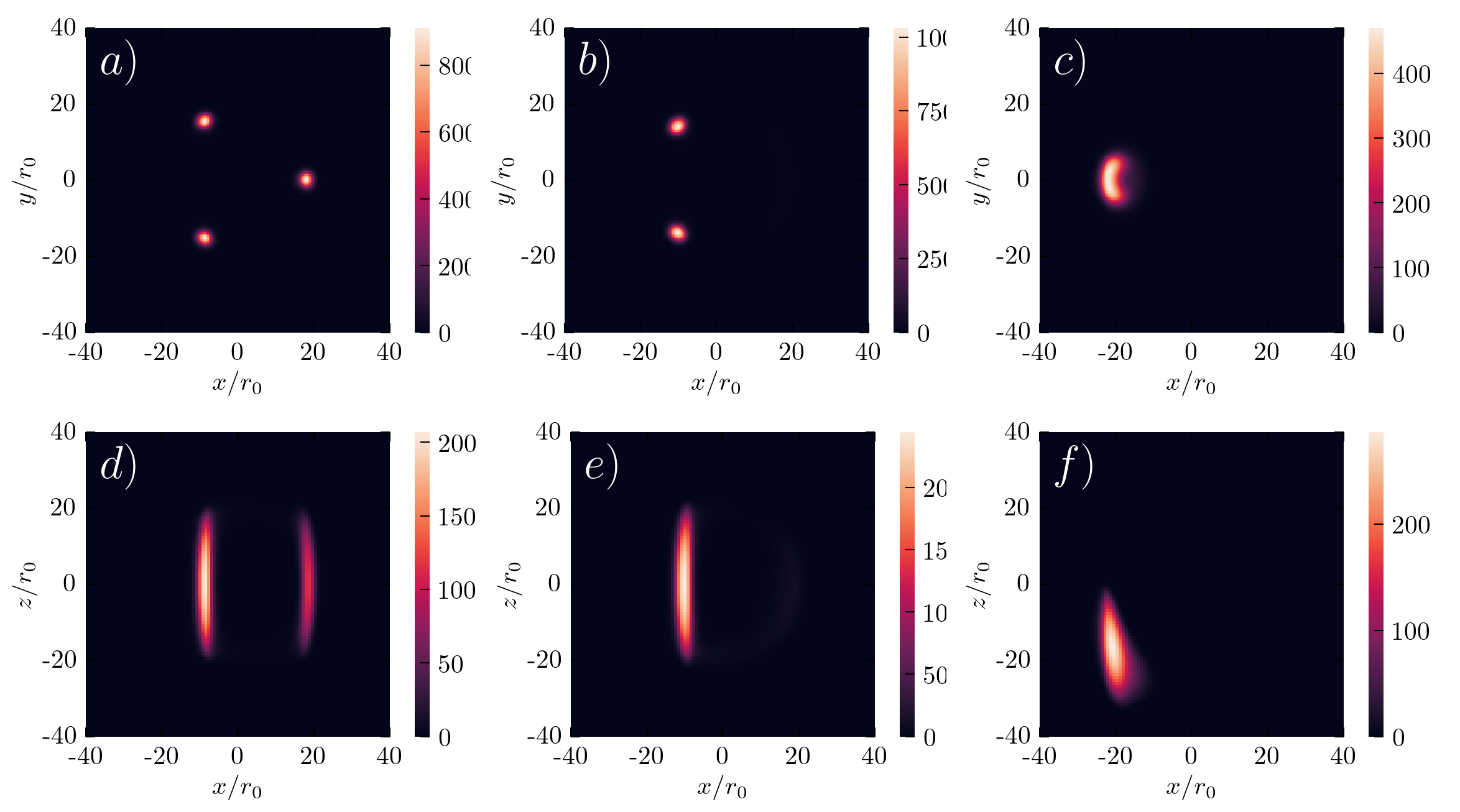}
\caption{Integrated densities $\rho(x,y)$ ((a)-(c)) and $\rho(x,z)$ ((d)-(f)) for a varying gravitational strength $mg = 0.0001 m g_{\rm E}$ ((a), (d)), $mg = 0.0025 m g_{\rm E}$ ((b), (e)) and $mg = 0.5 m g_{\rm E}$ ((c), (f)) and an angle $\theta_g = \pi/4$ (see Eq.~\ref{eqn:gravity}). The bubble trap parameters are the same as in Fig.~\ref{fig:fig2}. The colorbars indicate the value of $\rho(x,y)$ and $\rho(x,z)$ in units of $r_0^{-2}$. }
\label{fig:fig9}
\end{figure}

The structural diagram reported in Fig.~\ref{fig:fig2} has been computed assuming zero gravity conditions. Experimentally, and as stated before, microgravity conditions are achievable in the NASA Cold Atom laboratory in the International Space Station. However, it is interesting to explore the effect of a gravitational force on the dipolar arrangements that have been reported, to study how robust these structures are with respect to gravity. We account for gravitational effects through the inclusion of the one-body potential of Eq.~\ref{eqn:gravity}.
We have performed calculations for $\varepsilon_{\rm dd}= 1.47, N= 31500$ (which 
yields a solid state with 3 droplets in the absence of gravity) varying the 
strength of the gravitational field. We first consider a gravity vector with 
$\theta_g=0$. The results are shown in Fig.~\ref{fig:fig8}. For values of the 
gravity strength $mg < 0.03 m g_{\rm E}$, the structure of the dipolar BEC is 
not significantly altered, while for $mg > 0.04 m g_{\rm E}$ we observe the 
melting of the solid configuration into a superfluid, clusterless state. This 
indicates that, for these parameters, the gravitational force field of the Earth 
would destroy the solid arrangement of droplets, in contrast to what happens in 
the different parameter regime considered in Ref.~\cite{Ciardi2024}. This is 
because our calculations do not lie in the thin-shell limit considered in 
Ref.~\cite{Ciardi2024}, since a tighter trap confinement implies a higher energy 
cost for particles to accumulate in a reduced space at the bottom of the trap. 
We have also performed calculations changing the relative orientation between 
the gravity field and the $z$-axis to $\theta_g=\pi/4$. We 
show the results in Fig.~\ref{fig:fig9}. In this case, we find that the three 
droplet structure remains unaltered up to $mg \simeq 0.0025 m g_{\rm E}$, where 
gravity induces a transition into a two droplet state. Further increasing the 
gravitational strength leads to the merging of the two dipolar clusters into 
one.

\subsection{Comparison with previous works}

Previous works have explored the formation of supersolid structures in curved 
trapping geometries. Authors in Ref.~\cite{Ciardi2024} also explore the 
supersolid properties on dipolar BECs confined in a bubbled trap. However, they 
do so through the use of an ab-initio Monte Carlo method. In comparison to the 
results from Fig.~\ref{fig:fig2}, their results involve a considerably lower 
number of particles ($N<300$), a shell of smaller radius, and lie in the 
thin-shell limit, where Eq.~\ref{eqn:eGPE} is no longer valid, and a 
pseudopotential that accounts for the curvature of the confinement has to be 
applied. Ref.~\cite{Ciardi2024} shows  supersolid and 
solid structures with four dipolar clusters along the equator of the sphere, 
much like the structure that emerges in region 4 of the structural diagram of 
Fig.~\ref{fig:fig2}. The physics of dipolar BECs under a bubble trap
has also been studied in Ref.~\cite{ghosh2024}. In comparison to our work, they employ a
higher number of particles ($N=60000$), a considerably higher trap radius $R=21 
\mu$m and a detuning not equal to the Rabi coupling, $\Delta \neq \Omega$, which 
favours the emergence of a higher number of dipolar clusters compared to the 
structures reported in our Fig.~\ref{fig:fig2}. Remarkably, it is reported 
that supersolidity can be induced in solid arrangements of 
dipolar clusters by inducing a rotation in the system~\cite{ghosh2024}.
Other works have considered different kind of curved geometries, like toroidal 
traps~\cite{Tengstrand2021,Tengstrand2023} and box traps~\cite{Roccuzzo2022}. 
Toroidal traps produce ring shaped supersolids analogous to the ones found in 
this work, the difference being that under a shell-shaped confinement, and under 
microgravity conditions, the ring shape of the supersolid arises naturally from 
magnetostriction, instead of being 
imposed by the trapping confinement. In Ref.~\cite{Tengstrand2021}, the 
transition between a fluid ring into a supersolid, with 8 dipolar 
droplets as $\varepsilon_{\rm dd}$ increases, is reported. The authors employ a
toroidal trap with a considerably higher trapping strength of $\omega = 2 \pi 
\times 1000$ Hz compared to our parameters, which explains the higher number of 
clusters that they observe compared to our results, since a higher trapping 
confinement limits the length of the droplets along the polarization direction
an hence forces the system to organize in a higher number of clusters. On the 
other hand, the physics of anti-dipolar BECs have been 
recently addressed in toroidal traps~\cite{Mukherjee2024}, where the formation 
of stacks of ring-shaped droplets, which can coherently overlap to form a 
supersolid, has been reported.
In regards to results in a box potential, Ref.~\cite{Roccuzzo2022} explores 
different shapes for the box confinement, including a cylindrically shaped box 
trap, where ring solids and supersolids arise. Compared to our work, the authors 
in Ref.~\cite{Roccuzzo2022} consider a larger number of atoms, which induces 
the emergence of a larger number of droplets.

\section{\label{sec:conclusions}Conclusions}

We have studied the interplay between the anisotropy of the
dipole-dipole 
interaction and the curved geometry of the trapping potential for a dipolar 
condensate confined in a bubble trap. We have provided the structural diagram of a dipolar BEC as a function of the number of particles $N$ and the ratio between the
dipole length and the scattering length $\varepsilon_{\rm dd}$, and have reported
the emergence of a wide variety of ring shaped solid structures formed by 
arrangements of dipolar clusters along the equator of the trapping potential. We 
have characterized the transitions between different structures, showing that 
they are discontinuous, reminiscent of a first order phase transition in the 
thermodynamic limit. This establishes a clear connection between our system and 
a dipolar BEC trapped in a quasi-1D configuration, 
where, in the low density regime, a first-order phase transition
between a superfluid and a solid phase
takes place. We have also 
explored the high particle number limit ($N> 10^5$) and have observed that 
atoms accumulate in the central ring along the equator of the trap instead of 
forming secondary ring-like structures. We
have shown that supersolid states with varying number of dipolar clusters
can be engineered by changing $N$, $\epsilon_{\rm dd}$, and the effective radius 
of the trap, which is accomplished by tuning the detuning of the coupled rf 
field. In regards to the robustness of the dipolar structures, we have also considered the effect of a gravitational field and have shown that, for our parameters of choice, the gravitational field of the Earth would destroy an arrangement of dipolar droplets, forcing particles to accumulate at the bottom of the trap. Our results lead to the existence of ring shaped dipolar supersolid
states with varying number of clusters which, unlike in the case of ring-shaped traps, are not entirely forced by the confinement geometry, and arise instead as a result of the competition between the anisotropy of the DDI and the shell-shaped geometry of the bubble trap.

The study of the excitations of these ring
supersolids remains a relevant question to be addressed, since it could lead to 
an experimental protocol to probe the gas-to-supersolid transition by means of 
measuring excitation frequencies. Also, it remains an open question how finite 
temperature could affect the physics of the dipolar system under these trapping 
conditions. Recent results~\cite{Sohmen2021, Baena2023, Baena2024} reveal an important
impact of thermal fluctuations on dipolar gases, leading to the 
counterintuitive formation of a supersolid by heating in the ultracold regime.

\acknowledgments{
We acknowledge financial support from Ministerio de Ciencia e Innovaci\'on 
MCIN/AEI/10.13039/501100011033
(Spain) under Grant No. PID2020-113565GB-C21 and
from AGAUR-Generalitat de Catalunya Grant No. 2021-SGR-01411. 
J.S.B and R.B acknowledge European  Union-NextGenerationEU, Ministry of Universities and
Recovery, Transformation and Resilience Plan, through a call from the Technical
University of Catalonia. R.B. further acknowledges funding from the European Union’s Horizon 2020 research and innovation programme under the Marie Sklodowska-Curie Grant Agreement No. 101034379.}

\appendix

 \begin{figure}[t]
\centering
\includegraphics[width=\linewidth]{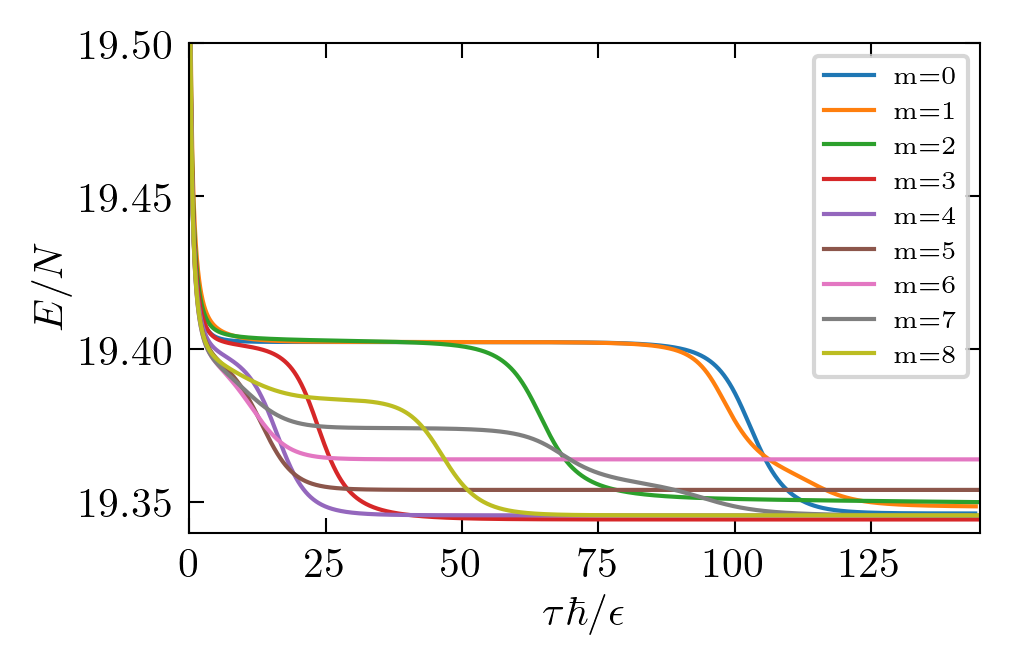}
\caption{Energy per particle as a function of the imaginary time for the different initial conditions given by Eq.~\ref{eqn:initial}. The parameters are $\varepsilon_{\rm dd}=1.47$, $N=31500$, $\Delta/\epsilon=400$}
\label{fig:fig_a}
\end{figure}

\section{\label{sec:numerics}Numerical implementation}

All the results shown in this work are obtained by propagating Eq.~\ref{eqn:eGPE} in imaginary time. To do so, we discretize space and work with a grid of $(N_x,N_y,N_z)=(200,200,100)$ points, in a simulation box of size $(L_x/r_0,L_y/r_0,L_z/r_0)=(80,80,80)$. During imaginary time propagation, we employ a time step of $d\tau \hbar / \epsilon = 0.0005$. The DDI term of Eq.~\ref{eqn:eGPE} is evaluated by computing its Fourier transform through an FFT routine. As mentioned in the main text, during the imaginary time evolution of Eq~\ref{eqn:eGPE} it is very likely for the system to get stuck in a metastable state. Because of this, we set multiple calculations starting from a variety of initial conditions when computing the ground state of the system for a given set of values $(N,\varepsilon_{\rm dd}),\Delta/\epsilon)$. The set of initial conditions employed is given by
\begin{align}
 \Psi_0({\bf r}) &= \exp \left( -\frac{\omega_0 \hbar}{\epsilon} \left( r/r_0 - \sqrt{\Delta/\epsilon} \right)^2 \right) \nonumber \\
 & \times \left( 1 + 0.5 \cos \left( m \phi \right) \right)
 \label{eqn:initial}
\end{align}
where $r = \abs{{\bf r}}$, $\phi$ is the azimuthal angle and $m = 0, 1, 2, ..., 8$. This allows us to start from initial configurations close in shape to those with $m$ number of clusters. After imaginary time evolution, we retain the state with the lowest energy as the ground state. In order to illustrate this process, and to showcase the rich metastable landscape of the system, we show in Fig.~\ref{fig:fig_a} the energy as a function of the imaginary time for the different initial states for $\varepsilon_{\rm dd}=1.47$, $N=31500$, $\Delta/\epsilon=400$.

\bibliography{refs}

\widetext

\end{document}